\documentclass[10pt,
preprint,
nofootinbib,
 amsmath,amssymb,
 pre,
longbibliography,
floatfix,
]{revtex4-1}
\usepackage{graphicx}
\usepackage{epstopdf} 
\usepackage{dcolumn}
\usepackage{bm}
\usepackage{hyperref}
\begin{document}
\title{Interface-mediated thermomechanical effects during high velocity impact between monocrystalline surfaces}
\author{Zhenqi Yang}
\affiliation{Department of Mechanical and Industrial Engineering, Northeastern University, Boston 02115}
\author{Sinan M\"{u}ft\"{u}}
\affiliation{Department of Mechanical and Industrial Engineering, Northeastern University, Boston 02115}
\author{Moneesh Upmanyu}
\email{mupmanyu@neu.edu}
\affiliation{Group for Simulation and Theory of Atomic-Scale Material Phenomena (stAMP), Department of Mechanical and Industrial Engineering, Northeastern University, Boston, Massachusetts 02115, USA}

\begin{abstract}
High velocity impact between crystalline surfaces is important for a range of material phenomena, yet a fundamental understanding of the effect of surface structure, energetics and kinetics on the underlying thermo-mechanical response remains elusive. Here, we employ non-equilibrium molecular dynamics (NEMD) simulations to describe the nanoscale dynamics of the high velocity impact between commensurate and incommensurate monocrystalline (001) copper surfaces. For impact velocities in the range 100-1200 m/s, the kinetic energy dissipation involves nucleation and emission of dislocation loops from defective sites within the rapidly forming interface, well below the bulk single-crystal yield point. At higher velocities, adiabatic dissipation occurs via plasticity-induced heating as the interface structurally melts following the impact. The adhesive strength of the reformed interface is controlled by the formation and nucleation of dislocations and point defects as they modify the interfacial energy relative to the deformed bulk. As confirmation, the excess interface energy decreases monotonically with increasing impact velocity. The relative crystal orientation of the surfaces equally important; the grain boundaries formed following incommensurate impact exhibit higher impact resistance, with smaller defect densities and interfacial enthalpies, suggesting an enhanced ability of the grain boundaries to absorb the non-equilibrium damage and therefore facilitate particle bonding. Our study highlights the key role played by the atomic-scale surface structure in determining the impact resistance and adhesion of crystalline surfaces.
\end{abstract}

\maketitle

\section{Introduction}
High strain-rate response of crystalline materials continues to be increasingly relevant for the processing and performance of advanced structural materials. Applications of note include both traditional and emerging technologies, such as thermal/cold spray technologies~\cite{swp:FauchairMontavon:2008, swp:KurodaKatanoda:2008}, shock sintering~\cite{book:Meyers:1994}, explosive welding~\cite{book:Lancaster:1999}, laser machining and ablation~\cite{book:Ready:1997, swp:MeyersWark:2003},  and terahertz radiative materials~\cite{swp:ReedJoannopoulos:2007}. The viability of these processes hinges on microstructural changes induced by high-velocity impingement of disparate materials. The impact resistance and subsequent adhesion is critical for the design of defect-engineered systems for a variety of structural applications, from superhard and irradiation-resistant coatings to joining of dissimilar materials. 

Current understanding of the mechanics and physics of the impact is based on experimental and computational studies of shockwave (SW) dynamics within bulk crystals~\cite{swp:AssadiKreye:2003, swp:KlinkovRein:2005, swp:SchmidtKreye:2006}. Flyer-plate impact and related techniques have proven useful in developing a qualitative understanding of the macro-scale evolution of shocked metal microstructures~\cite{book:Meyers:1994}. It is now well-known that in low stacking fault metals such as copper, SW-induced plasticity is associated with two distinct regimes: (i) slip, dominated by bulk-nucleated dislocations, and (ii) strain accommodation by nucleation of planar defects such as stacking faults and twins~\cite{plasticity:Edington:1969}. The experiments have also highlighted the effect of loading~\cite{plasticity:Edington:1969, swp:GrayFrantz:1989, plasticity:LiaoGunderov:2004}, grain size~\cite{swp:ZerilliArmstrong:1988}, and crystal orientation~\cite{swp:CaoMeyers:2010, swp:MeyersWark:2003} on the nature of the dislocated microstructure and its transition to slip-twinning. 
\begin{figure}
\begin{center}
\includegraphics[width=0.7\columnwidth]{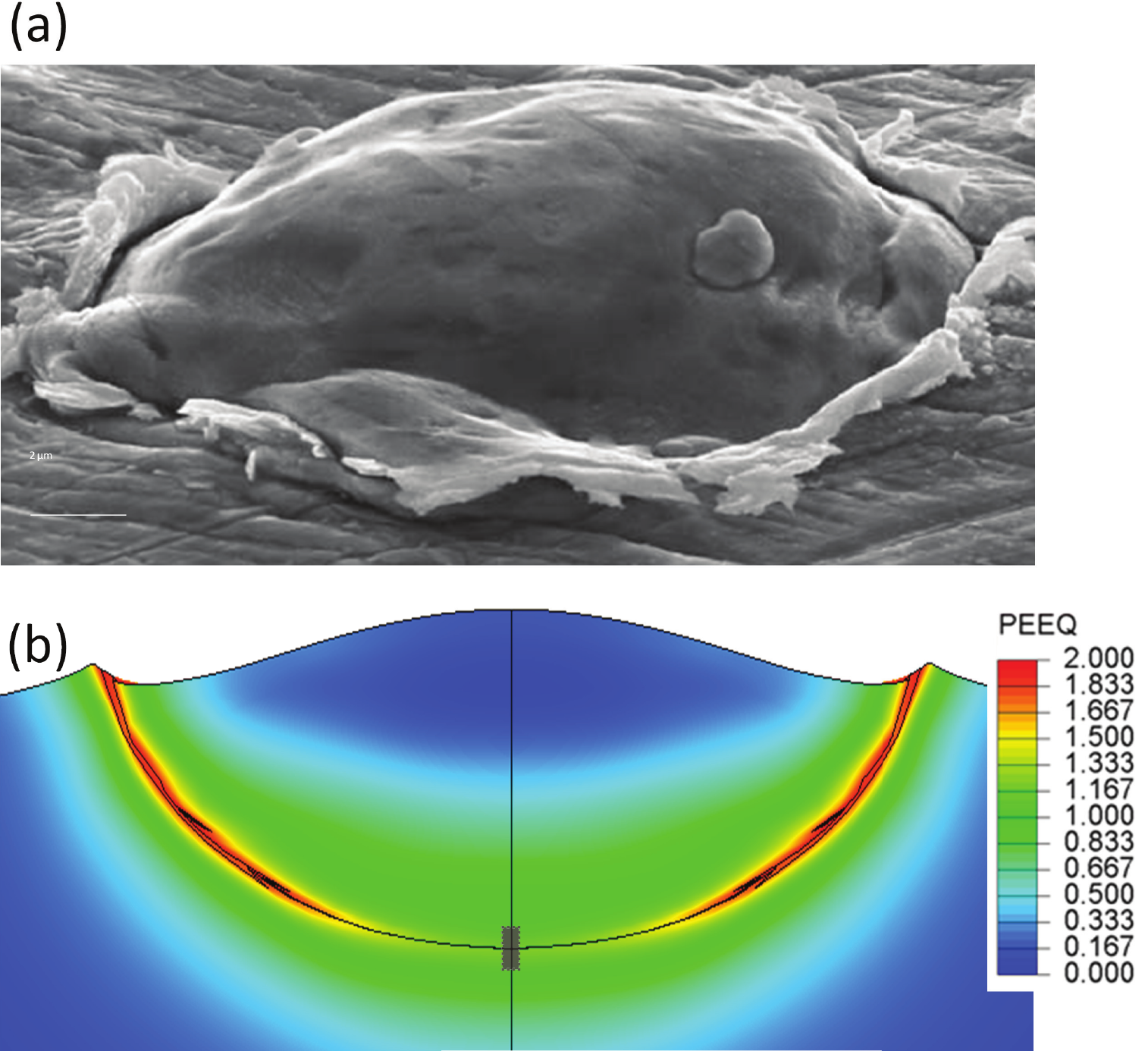}%
\caption{Comparison of particle-substrate morphologies observed in (a) experiments and (b) continuum-scale simulations for Cu-on-Cu particle impact at velocity $u_p=300$\,ms$^{-1}$. Experimental and computational images are taken from Refs.~\cite{swp:AssadiKreye:2003} and~\cite{swp:YildirimMuftu:2012}, respectively. The color in the computational image corresponds to the variation in equivalent plastic strain (PEEQ). The dashed rectangular box a schematic illustration of the approximate interfacial region simulated in this study.}%
\label{fig:figure1}
\end{center}
\end{figure}
Complementary computational approaches, in particular atomic-scale simulations, are well-suited for studying the early stage response. Indeed, simulations of SW dynamics in metals have uncovered important microscale mechanisms. In one such pioneering study, Holian and Lomdahl~\cite{swp:HolianLomdahl:1998} employed non-equilibrium molecular dynamics (NEMD) simulations to identify deformation mechanisms associated with planar and non-planar SW fronts. Subsequent efforts have quantified the Rankine-Hugoniot relations that characterize the dynamics of SWs with varying crystal orientations, and their effect on defect nucleation and evolution~\cite{swp:BringaCaturla:2004, swp:CaoBringaMeyers:2007, swp:TanguyRavelo:2003}.\vspace{6pt}

Although the combination of experimental observations and molecular/atomistic simulations have provided fundamental insight on the strain accommodation mechanisms, connections between the two have been tenuous as the atomistic frameworks are limited in the accessible length- and time-scales. In contrast, continuum-scale studies such as finite-element method (FEM) based approaches have the capability to make direct contact with experiments. They provide a good description of the overall energy balance of the particles (see Fig.~\ref{fig:figure1} for a direct comparison), and recent efforts have revealed some of the key aspects of the dissipation processes during high strain-rate plasticity~\cite{swp:AssadiKreye:2003, swp:SchmidtKreye:2006, swp:KurodaKatanoda:2008, swp:YildirimMuftuGouldstone:2011, swp:YildirimMuftu:2012}. Nevertheless, these approaches fail to capture the metallurgical and physical interactions between the impacting free surfaces that themselves are set at the atomic-scale. They influence microstructural changes following the impact~\cite{swp:BalaniKarthikeyan:2005, swp:BorchersKreye:2003, swp:BorchersKreye:2005, swp:KimKuroda:2008, swp:KingJahedi:2009}, including but not limited to localized melting and recrystallization, thermal gradients and their evolution, dislocation nucleation and dynamics, interfacial diffusion and bonding, etc. Evidently, a fundamental understanding of these interrelated surface-mediated phenomena requires access to atomistic scales, the principal focus in this study.

The importance of the crystalline surfaces cannot be overstated; they modify the response as they themselves act as efficient sources and sinks for defects - dislocations, vacancies, impurities - during high velocity impact. Although this has been theorized in the context of several joining and deposition techniques, past NEMD-based studies on shockwave dynamics have been limited to the bulk response. The role of surfaces on energetic impact has been explored in the context ion beam deposition, an analogous processing technique that is used to tailor surface properties of as-synthesized thin films. It often involves high energy impingement of molecular clusters onto the growing film, not unlike particle impact in spray processes~\cite{ionbeam:ClevelandLandman:1992, ionbeam:HaberlandMoseler:1995, ionbeam:Popok:2011}. Dissipation of the kinetic energy leads to shock wave generation from the free surface following cluster impact, and the ensuing phenomena results in defect generation~\cite{ionbeam:JuWeng:2002}, surface reconstructions~\cite{ionbeam:RongwuYunkun:1996}, subsurface interstitial and void formation~\cite{ionbeam:JuWeng:2002, ionbeam:RongwuYunkun:1996}, and longer time-scale surface diffusion and morphological evolution~\cite{ionbeam:Chason:1997, ionbeam:NorrisAziz:2011}, each of which is sensitive to the atomistics of the free surface. Atomic-scale simulations have shown that these events modify variations in local density, stresses, temperature and deformation following impact, all of which together determine the nature and extent of cluster adhesion and thin film evolution~\cite{ionbeam:ClevelandLandman:1992, ionbeam:HaberlandMoseler:1995, swp:ValentiniDumitrica:2007}. Of particular relevance is a recent all-atom study by Anders et al.~\cite{swp:AndersUrbassek:2012} involving multi-million atom Cu clusters impinging on crystalline copper films. Hypervelocity impact ($>4000$\,m/s) of nanoscale clusters show that the transition from nanoscale to micron-scale impact is in large part controlled by increasing surface-mediated plastic deformation and crystallization.

In this study, we self-consistently investigate the role of surfaces and interfaces on particle impact by systematically studying the low  (room) temperature high velocity impact of two crystalline copper plates using NEMD simulations. The plates are chemically pure Cu single crystals that terminate at atomically flat and relaxed surfaces. Although this appears to be a highly idealized scenario, it allows us to identify nanometer-scale thermo-mechanical phenomena near the interface at sub-nanoseconds time-scales (e.~g. boxed region in Fig.~\ref{fig:figure1}). Since the extent of the plates is of the order of a few tens of nanometers, the surface impact corresponds to a particle-substrate system whose size and roughness are micron-scale or larger.  The bulk away from the surfaces is devoid of pre-existing defects; this again relates to annealed crystals with low defect densities. The simulations are performed for  commensurate and incommensurate (relatively twisted) \{001\} surfaces at varying relative velocities, and in each case we report the salient aspects of the underlying microstructural evolution.

The rest of the article is organized as follows: we describe the details of the NEMD simulations, in particular the simulation geometry and the modified boundary conditions that prevent the SWs from reflecting and refocusing back to the interface. The results for varying impact velocities and degree of incommensurability are presented as analyses of the energetics, kinetics and thermomechanical properties of the crystals at and near the interface. They include evolution of locally deformed structure, gradients in temperature, normal virial and resolved shear stresses, and interface enthalpies. We discuss the relevance of our results on particle adhesion during low temperature particulate consolidation, in particular cold spray.

\section{Computational Methodology}
\subsection{Geometry}
The simulation cell is shown in Fig.~\ref{fig:figure1}. It consists of a three-dimensional copper bicrystal system composed of two identically-sized monocrystalline plates. They are modeled as infinite half-spaces along the $z$-axis by applying in-plane periodic boundary conditions (PBCs). The atomically flat surfaces are normal to the $\langle001\rangle$ family of crystal orientations ($z-$axis).  The opposing non-impacting ends are free boundaries that bound the shockwave absorbing layers (SWALs, yellow-colored atoms in Fig.~\ref{fig:figure1}), as described later in the section.
\begin{figure}[htp]
\begin{center}
\includegraphics[width=\columnwidth]{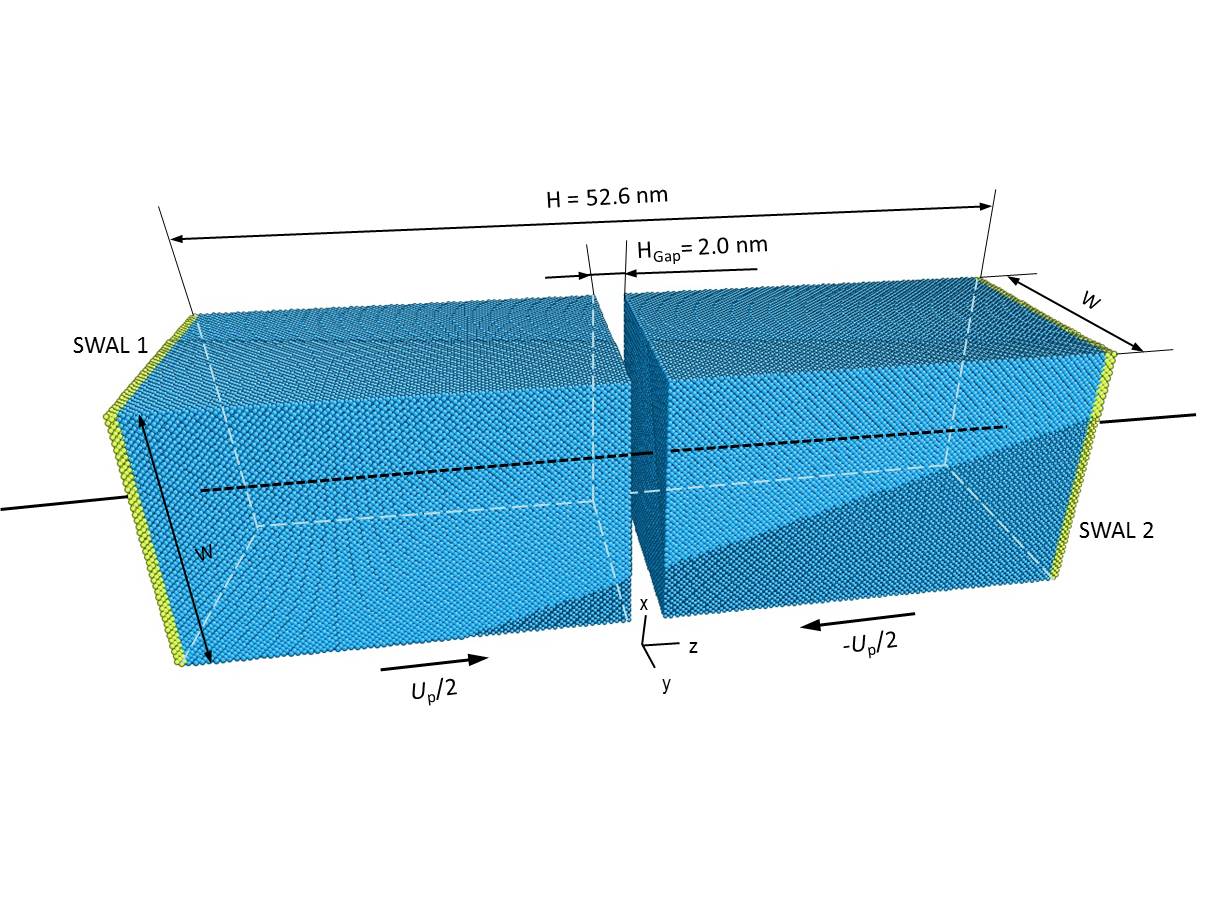}%
\caption{The computational cell employed for non-equilibrium molecular dynamics simulations of impact between two copper plates. The width of each monocrystalline plate is $w=11-15$\,nm. The boundary atoms along the impact direction (colored yellow) are free surfaces that serve as shockwave absorbing layers (SWALs), see text for details.}%
\label{fig:figure2}
\end{center}
\end{figure} 

The degree of commensuration at impact is prescribed by rotating the two crystals relative to each other symmetrically about the $z-$axis. Denoting this change in orientation of the two crystals as $\pm\theta/2$, the two crystals then form a $\langle 001\rangle$ twist grain boundary (TGB) with a misorientation angle $\theta$. In addition to the fully commensurate case $\theta=0$, we explore the effect of three misorientation angles, $\theta=15^\circ$, $30^\circ$, and $45^\circ$.  The commensurate bi-plate system consists of 990,000 atoms with cell dimensions $15\times15\times50$\,nm$^3$ [$42\times42\times140$ face-centered cubic (FCC) unit cells]. The in-plane dimensions of the incommensurate cells are different to ensure that the cell is periodic. The impact area of the plates ranges from $125$ to $210$\,nm$^2$, and the cell contains $500,000$-$900,000$ atoms. In each case, the size of the simulation cell is sufficiently large such that periodic and free boundaries have a minimal effect on the thermo-mechanical phenomena at the interface~\cite{swp:TanguyRavelo:2003}. 

\subsection{Molecular dynamics simulations}
The MD simulations are performed using the Large-scale Atomic/Molecular Massively Parallel Simulator (LAMMPS) software package~\cite{md:Plimpton:1995}. The embedded-atom-method (EAM) potential for copper is used to describe the interatomic interactions~\cite{intpot:Mishin:2001}. The empirical potential has been used widely to describe a range of thermo-mechanical phenomena in copper, including high strain-rate deformation during flyer plate and cluster impact studies~\cite{swp:BringaCaturla:2004, swp:AndersUrbassek:2012, swp:LuoAn:2010}. 

\subsubsection{Initial relaxation}
The two crystals are first equilibrated under ambient conditions, i.e. room temperature ($T_0=293$\,K) and pressure of $P=0$\,bar. Equilibrium isothermal-isobaric (NPT) MD simulations are performed on the as-constructed crystals (FCC lattice parameter $a=3.61$\,{\rm \AA}) using a time step of $1$\,fs, and Nos\'{e}-Hoover thermostat and barostat~\cite{atsim:Nose:1984, atsim:Hoover:1985, book:AllenTildesley:1989}. The entire cell equilibrates within $50$\,ps. The atomic trajectories are then gradually modified to an NVE ensemble via two short equilibrium MD simulations performed in series:  i) a $25$\,ps relaxation within a canonical MD simulation at $T=293$\,K by simple velocity rescaling every $100$ time steps~\cite{book:AllenTildesley:1989}, ii) followed by an additional $25$\,ps simulation in the microcanonical NVE ensemble without any temperature control. The bi-plate system equilibrates to an average temperature $T=293.4$\,K with minimal residual stress.

\subsubsection{NEMD}
The equilibrated plates are placed next to each other such that the impinging free surfaces are separated by a small gap $H_{gap}=2.0$\,nm greater than twice the cut-off radius of the short-ranged interatomic potential, i.e. the two crystals are initially non-interacting (Fig.~\ref{fig:figure1}). The impact is initiated by assigning equal and opposite initial velocities to the center of mass of each crystal normal to the impinging surfaces ($z-$axis), $\pm u_p/2$. The relative impact velocity $u_p$ is varied from $100-1200$\,m/s in $100$\,m/s increments. The atomic trajectories are generated using NVE MD simulations with a larger time step, $\Delta t=2$\,fs. Each impact simulation is performed for $50$\,ps.

\subsubsection{Shockwave absorbing layers (SWALs)}
On impact, a pair of SWs are produced that travel away from the interface formed at  $z=0$. The compressed region bounded by the SWs evolves with piston velocities $\mp u_p/2$ such that it has no net velocity~\cite{swp:BringaCaturla:2004}. The SWs eventually impinge on the free boundaries and can reflect back towards the interface. Since we are interested in the response of semi-infinite cystals over the MD time-scales, the reflected SW is an artifact that modifies the microstructural evolution. Its effect is usually dramatic as it often leads to tensile damage and ultimately interfacial decohesion~\cite{swp:LuoAn:2010, swp:JarmakaniMeyers:2008}. 
\begin{figure}[htbp]
\begin{center}
\includegraphics[width=0.8\columnwidth]{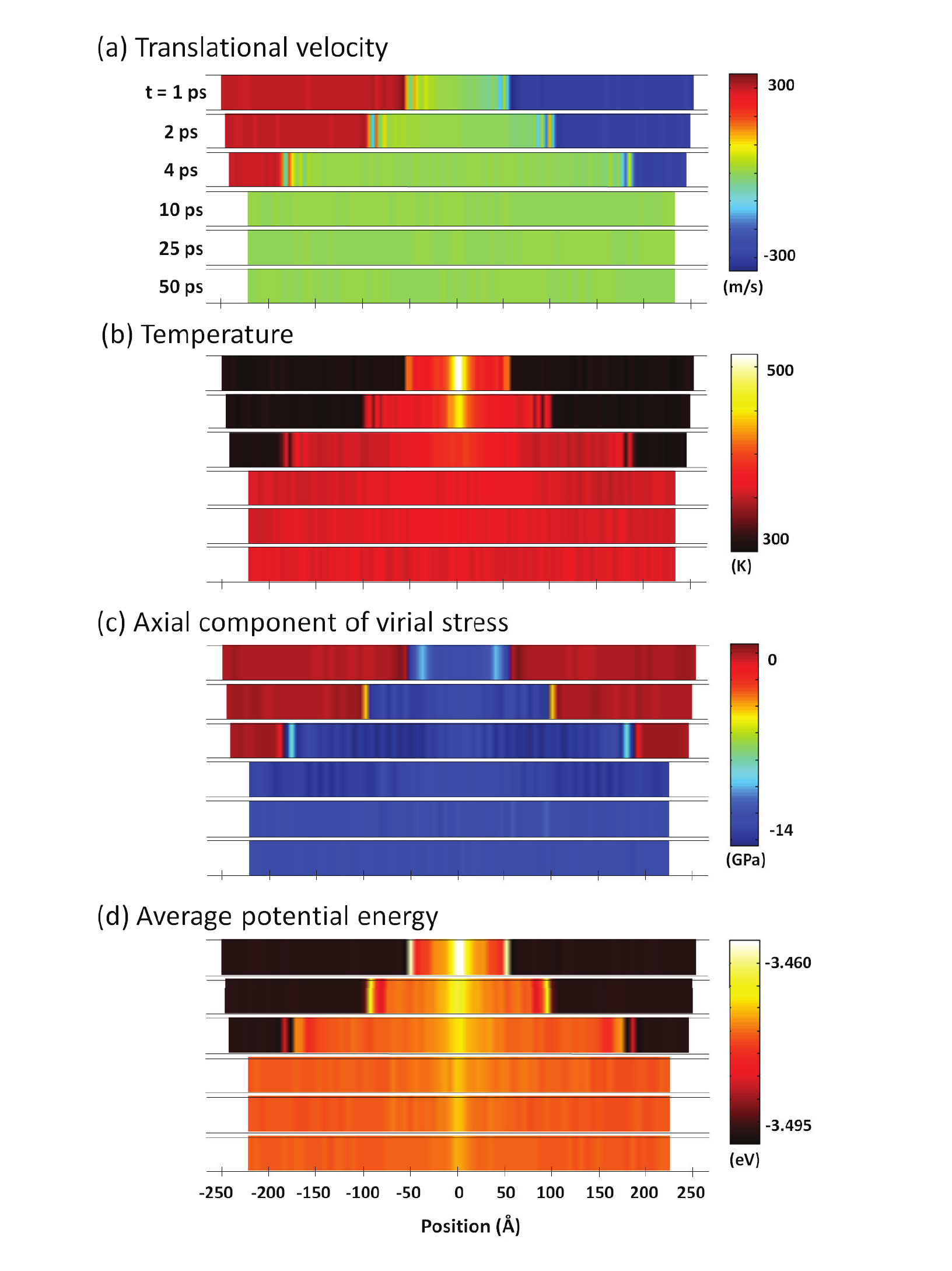}%
\caption{1D contour plots of (a) translational velocity, $u_z$ (b) temperature, $T$ (c) component of virial stress $\sigma_{zz}$, and (d) average potential energy, $\bar{E}$, for commensurate impact ($u_p = 600$\,m/s) at six representative instants. The SWALs are applied to the simulation cell at $t=5.32$\,ps. The interface formed post-impact is centered at $z = 0$\,{\rm \AA}.}%
\label{fig:figure3}
\end{center}
\end{figure}
Several strategies have been put forth to absorb SWs at appropriate simulation cell boundaries. They involve viscous damping, controlling the velocity of the boundary atoms in an artificial piston-like manner that sustains the shocked and compressed state, and energetic absorption by separate NVE integration on the boundary layers~\cite{swp:ZhaoStrachan:2006, swp:BolestaThompson:2007, swp:CawkwellThompson:2008}.  Here, we employ two $0.5$\,nm thick SWALs assembled on the non-impacting free boundaries (Fig.~\ref{fig:figure1}). As the SW reaches the SWALs, the velocities of the SWAL atoms are frozen out, and then the NVE integration is resumed. The time $t_c$ at which the SWs reach the SWALs is critical for our implementation, and it is calculated dynamically as the time at which the average longitudinal velocity of the SWAL atoms becomes zero. The finite width of the SW envelope can lead to errors in $t_c$, but for the copper interatomic potentials used here the front is relatively sharp with a thickness as low as $\sim0.3$\,nm. Figure~\ref{fig:figure3} shows the evolution of the bi-plate system with $t_c=5.32$\,ps. It is clear that the SWs are completely absorbed by the SWALs and therefore do not modify the subsequent evolution of the shocked crystals.

\subsection{Thermo-mechanical characterization}
The evolution of local quantities of interest such as temperature and stresses are monitored by discretizing the simulation cell into $0.5-0.8$\,nm thick 1D bins along the $z$-axis and then analyzing the bin-averaged properties~\cite{swp:LuoAn:2010}. Contributions due to the translational motion of the mass-center of each bin are ignored. The volume-averaged atomic virial stress is extracted as~\cite{nw:DiaoGallDunn:2004} 
\begin{equation}
\label{eq:STR}
\sigma_{\alpha\beta}=\frac{1}{V}\left[ mv_{i\alpha}v_{j\beta} + (f_{ij})_{\alpha}(r_{ij})_{\beta} \right],
\end{equation}
where $(\alpha\beta)\equiv (x, y, z)$, $V$ is the simulation cell volume, $m_i$ and $v_i$ are the atomic mass and velocity, and $f_{ij}$ is the force exerted by each $j^{\rm th}$ neighbor located at a distance $r_{ij}$. The summation over $i$ runs over all the atoms in each bin, and that over $j$ is over all neighbors within the cut-off distance associated with the short-range interaction potential. The components of the stress tensor (Eq.~\ref{eq:STR}) are used to calculate the local pressure and the resolved shear. The latter, which triggers the glide of the nucleated dislocations along slip planes, is calculated by transferring the local stress state from uniaxial compression to maximum shear~\cite{swp:HolianLomdahl:1998},
\[
2\tau=\sigma_{zz} - (\sigma_{xx} + \sigma_{yy})/2\,.
\]
As an illustration, Fig.~\ref{fig:figure3} shows the temporal evolution of the translational velocity $u_z$, average temperature $T$, virial stress $\sigma_{ij}$, and average potential energy $\bar{E}$ following the impact of two commensurate crystals at a velocity $u_p=600$\,m/s. The first three instants show SW fronts, evident from the sharp gradients in each of these quantities. The SW move away from the formed interface and are completely absorbed by the SWALs. 

The impact results in transient metastable phases and defects. The structural evolution is monitored using the common neighbor analysis (CNA) and the centro-symmetry parameter (CSP) that can distinguish between different crystal structures (FCC, BCC, HCP, disordered atoms), as well as point and extended defects such as vacancies, voids, dislocations and stacking faults~\cite{atsim:HoneycuttAndersen:1987, atsim:TsuzukiRino:2007}. 

\subsubsection{Interfacial energetics}
The atomic trajectories yield quantitative measures of non-equilibrium surface and interface energies $\gamma^H_s$ and $\gamma_i^H$ that together determine the bonding strength. Of particular interest is the post-impact evolution of the interface. Here and elsewhere, the interface is defined to be a region centered at  $z=0$ with thickness twice the lattice parameter, $w=2a\approx1$\,nm. Although the size of the region expands along its normal during the course of the simulation, we expect the qualitative trends to be independent of the interface thickness.

The energy of the interface, defined as the excess interaction energy of the bicrystal $E_t$ compared to perfect bulk copper crystal, is readily available from the MD simulations\footnote{The interfacial energetics is extracted based on the interaction energies between the particles readily available from the atomic-scale computations. The entropic contribution can be important but usually limited to equilibrium interfaces~\cite{fec:HoytUpmanyu:2010}, and its extraction is beyond the scope of this study as the interface structures here are necessarily non-equilibrium due to the impact-related processes. Since the entropic contributions are ignored, the extracted quantities is the interfacial enthalpy.}.  This definition ignores the effect of the deformed bulk away from the interface. We can correct for this by averaging the bulk energy over the deformed crystals sans the interfacial region $E_b$, i.e.
\begin{equation}
\label{eq:intEnergy}
\gamma_i^H = \frac{E_t-E_b}{A}.
\end{equation}
The interface energy so defined can be size dependent as the bulk energy $E_b$ changes with the size of the deforming bulk region over which it is averaged. Its temporal evolution is the quantity of interest as it captures the effect of increasing deformation at a constant (simulation cell) size. 

In order to directly capture the effect of the non-equilibrium thermo-mechanical state of surrounding bulk, we extract the energy difference associated with artificially cleaving the interface at $z=0$ into free surfaces. The deformed structure of the bulk is preserved by arresting the motion of the atoms away from the interfacial zone during the cleaving process. Then, the work done in cleaving yields the energy of the metallurgical bond, or the work of adhesion,  
\begin{equation}
\label{eq:WOA}
W^H = \frac{E_{s1} + E_{s2} - E_t}{A},
\end{equation}
where $E_t$ is the energy of the bicrystal, and $E_{s1}+E_{s2}$ is the energy of the system after cleaving the two crystals beyond the interaction range of the interatomic potential.

In addition to the energies, we also monitor thermomechanical properties of the interface. These include interface temperature $T_i$, and normal and maximum shear stresses $\sigma_{i}$ and $\tau_i$  within this region, respectively~\cite{swp:HolianLomdahl:1998}. 

\section{Results and Discussion}
\subsection{Energy dissipation mechanisms}
The impact triggers a variety of thermo-mechanical phenomena. In this section, we report salient aspects of the underlying atomic-scale mechanisms. Quantitative analyses are presented in Section~\ref{sec:tmanalysis}. We first report the results for commensurate impact, and then highlight differences directly attributable to the misorientation between the crystals during incommensurate impact.
\begin{figure}
\centering
\includegraphics[width=\columnwidth]{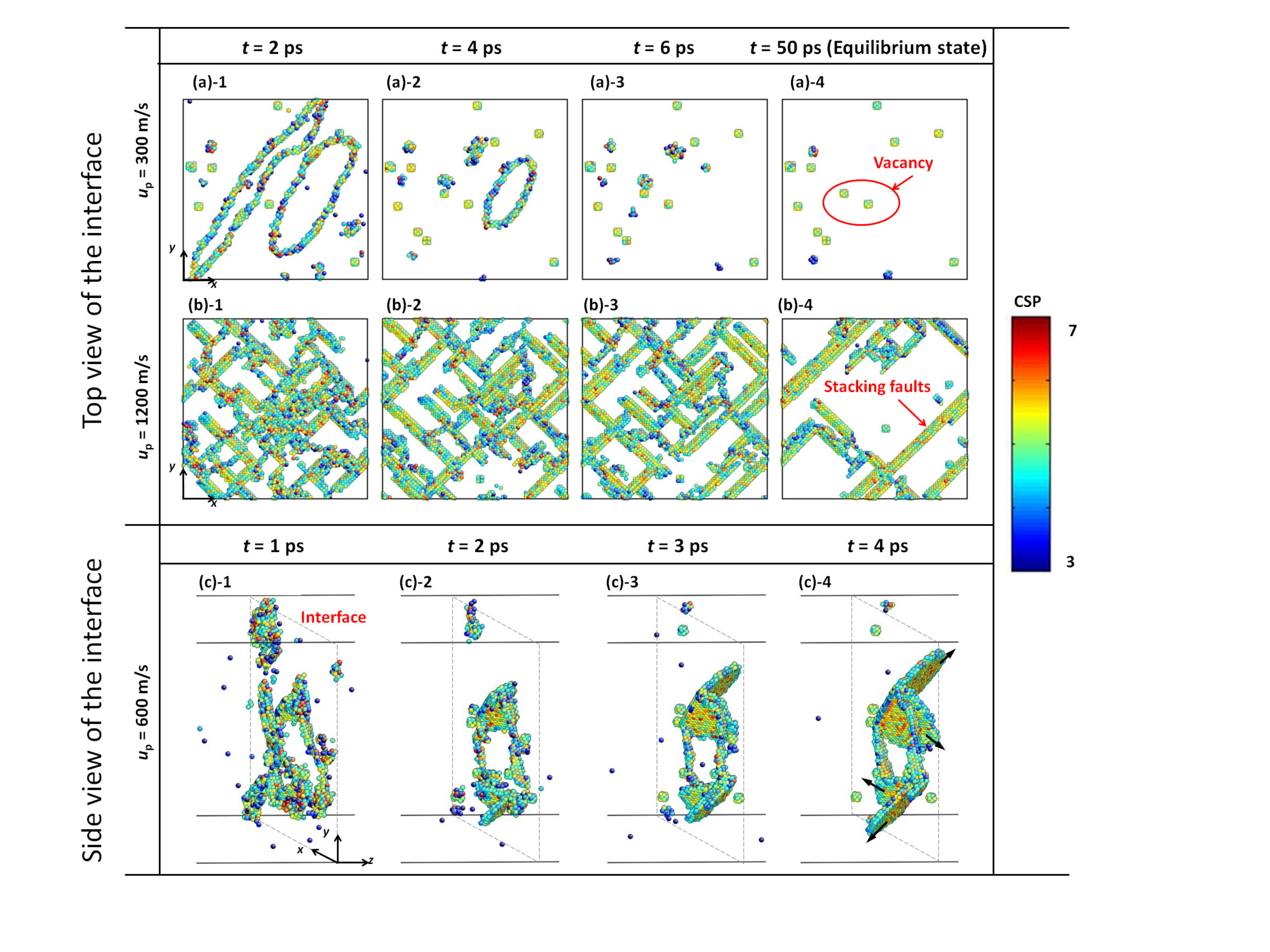}%
\caption{Atomic configuration of the interfacial region following impact between commensurate surfaces for different impact velocities, as indicated. Individual atoms are colored according to their local centro- symmetry parameter, and only defective atoms are shown. Snapshots of the top view (top two rows) are confined to the interfacial region, as defined in the text.}
\label{fig:figure4}
\end{figure}

\subsubsection{Commensurate impact} 
Figure~\ref{fig:figure4} shows the atomic configurations in the vicinity of the interface for three different impact velocities, $u_p=300$, $600$, and $1200$\,m/s. The centro-symmetry parameter is used to monitor the structural evolution~\cite{swp:JarmakaniMeyers:2008}. For clarity, perfect FCC atoms are removed from the configurations. In all cases, the kinetic energy is insufficient to induce surface melting yet the plates adhere to each other after impact. The initial dissipative processes occur over a few picoseconds and are mainly confined to the interface. For $u_p=300$\,m/s, we observe the formation of several vacancies and self-interstitials within this region (Fig.~\ref{fig:figure3}a). The disorder is a combined result of inhomogeneous displacements at the surfaces, and from the compressed trailing ends of the two SWs that start to move towards the SWALs following impact~\cite{swp:TanguyRavelo:2003}. The attendant rise in local temperature increases the diffusivity of these defects and they begin to cluster together. In some instances, they assemble into defect (vacancy) loops, apparent in the configuration at $t=2$\,ps. Subsequent structural rearrangement leads to a noticeable decrease in the non-equilibrium defect density as they anneal out via diffusion. In cases where vacancy loops are formed, the loop radius on average decreases as they begin to collapse under their own curvature. At a relatively longer time of $t=50$\,ps, the interfacial region is mostly commensurate and consists of isolated vacancies that eventually diffuse into the bulk at much longer times.

Increasing the impact velocity enhances the disorder at the interface and beyond a critical velocity, we see a qualitatively different phenomenon. As an extreme case, the interface formed following a $u_p=1200$\,m/s impact is severely disordered (Fig.~\ref{fig:figure4}b) due to the large uniaxial compression that occurs in the wake of the moving SW fronts. Although the average temperature does not reach the melting point, local amorphization at the interface is evident in the $t=2$\,ps configuration. The recovery is rapid and takes place via nucleation and growth of high pressure $\frac{1}{6}\langle112\rangle$ Shockley partials and their loops at multiple disordered sites at the interface. The loops grow away from the interface into the crystals (not shown) and this additional dissipation occurs at the expense of the non-equilibrium vacancy concentration at the interface. In some instances, the vacancy clusters migrate collectively and collapse along $\{111\}$ planes, leading to the nucleation of the dislocation loops~\cite{book:HirthLothe:1968, swp:BorchersKreye:2005}. As such, the interface region is slowly decorated by a cellular pattern of growing Shockley partials. At large velocities such as the case here, the dislocation loops are completely absorbed by the adjoining crystals~\cite{swp:HolianLomdahl:1998, swp:TanguyRavelo:2003}. Some of the partials intersect and re-orient to release some of the shear stresses required for their growth, resulting in the formation of prismatic loops that enhance the efficiency of the absorption~\cite{swp:TanguyRavelo:2003}. 

The response at intermediate impact velocities sheds light on the interface mediated plasticity. As an example, Fig.~\ref{fig:figure4}c shows response for $u_p=600$\,m/s impact. The $t=1$\,ps configuration consists primarily of locally disordered regions and vacancy loops prior to dislocation nucleation; the plasticity is negligible. As the structure recovers, the Shockley partial dislocations are nucleated nearly simultaneously from the remnant loop into the abutting crystals. Two sets of $\{111\}$ family of slip planes are activated and the leading partials of the pairs of dislocations grow away from the disordered region at interface, as indicated in the figure. Increasing the impact velocity leads to activation of all four equal slip planes, and at lower velocities the leading partials do not survive as they are unable to overcome the energy barrier for formation of a complete dislocation loop; they collapse back as vacancy clusters. It follows then that the plasticity is incipient in that it occurs past a critical impact velocity, $u_p^\ast\approx600$m/s.
\begin{figure}
\centering
\includegraphics[width=0.65\columnwidth]{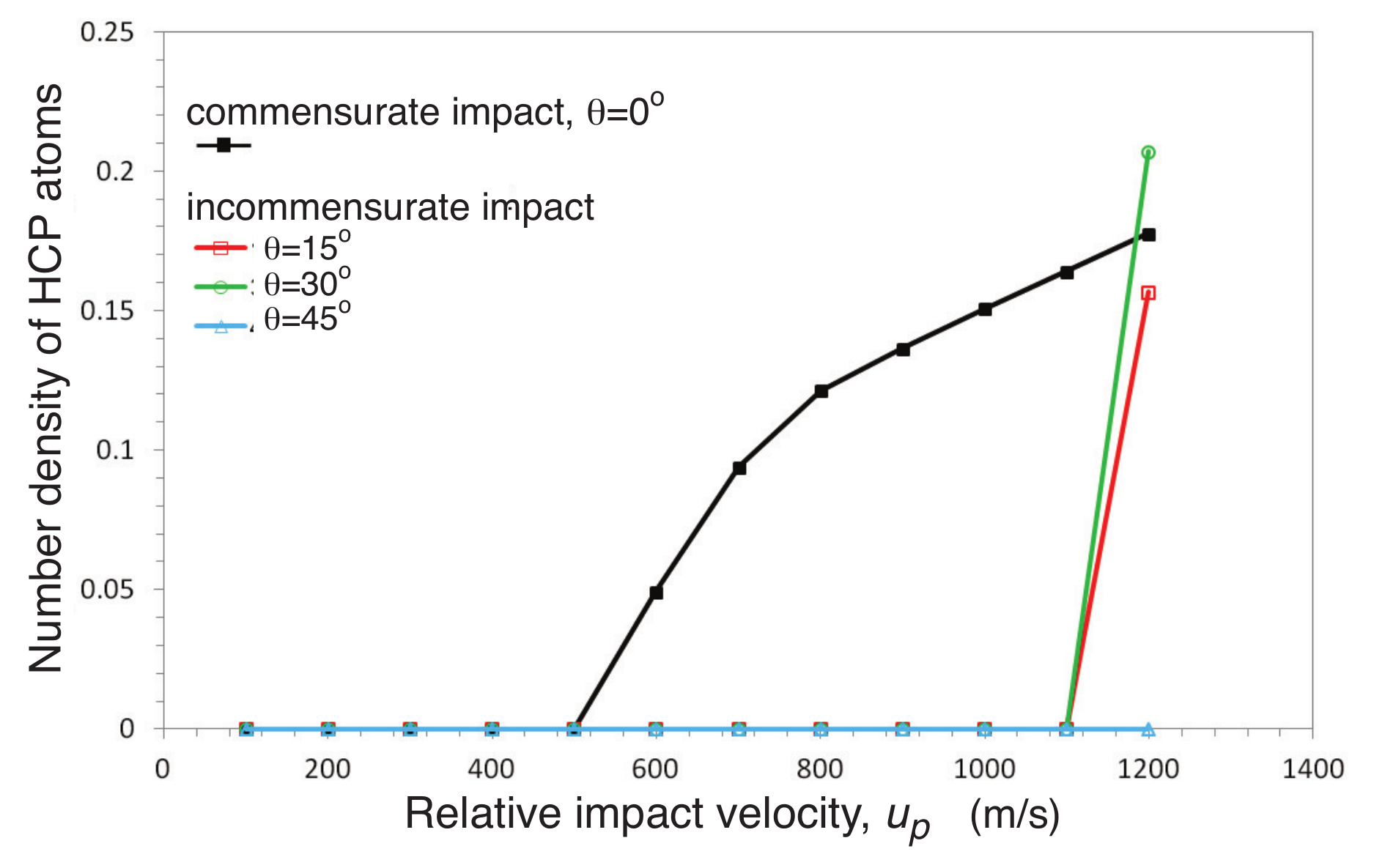}%
\caption{The HCP atoms normalized by total atoms in the system as a function of impact velocity. Each data point represents an average value of $2$\,ps relaxation after $50$\,ps simulation.}%
\label{fig:figure5}
\end{figure}

The deformation after the shock wave-front passes is driven by the roughly uniaxial compressed state in the wake of the shock-wave and around the interface region. In order to quantify the effect of surfaces on the incipient plasticity, we extract the precursor shear stresses. The resolved shear strength associated with activation of the slip planes is approximately $\tau_{CRSS}\approx3.2$\,GPa, which is comparable but a little higher than the theoretically calculated ideal shear strength for perfect single crystal copper, $\tau_{ISS}=2.6-3.1$\,GPa~\cite{swp:OgataYip:2002}. 
The yield strength extracted from the simulations at this critical impact velocity is $\sigma_Y\approx11$\, GPa. This is well below the bulk Hugoniot elastic limit (HEL) of $\sim32$\, GPa along the [001] direction reported in past atomic shockwave studies, for both Lennard-Jones (LJ) and EAM monocrystalline copper~\cite{swp:BringaCaturla:2004, swp:CaoBringaMeyers:2007}. Ignoring strain-rate hardening effects, the comparison indicates that the surfaces subvert the classical thermal fluctuation driven nucleation in shocked metals~\cite{swp:BringaCaturla:2004, swp:CaoBringaMeyers:2007, swp:TanguyRavelo:2003}. The highly non-equilibrium, disordered interfacial structure serves as a precursor for the defect generation and dislocation emission, fundamentally altering the threshold for incipient plasticity. 

In order to quantify the extent of plastic deformation, we have monitored the HCP atoms generated within the bicrystal. The temporal evolution of their number density, extracted within a 2\,ps equilibrium simulation following the 50\,ps data run, is plotted in Fig.~\ref{fig:figure5}.  As expected, their formation is coincident with the yield point associated with the critical impact velocity, $u_p^\ast$. For $u_p\ge u_p^\ast$, the number density initially increases with increasing impact velocity, a combined effect of the impact induced deformation and the shock-wave induced compression. Well above the critical value, though, the increase begins to saturate, consistent with the observation of high density of leading Shockley partials that begin to intersect as they grow away from the interface. The ensuing dislocation reactions on average relieve some of the plastic strain energy via formation of sessile partials.

\begin{figure}
\begin{center}
\includegraphics[width=\columnwidth]{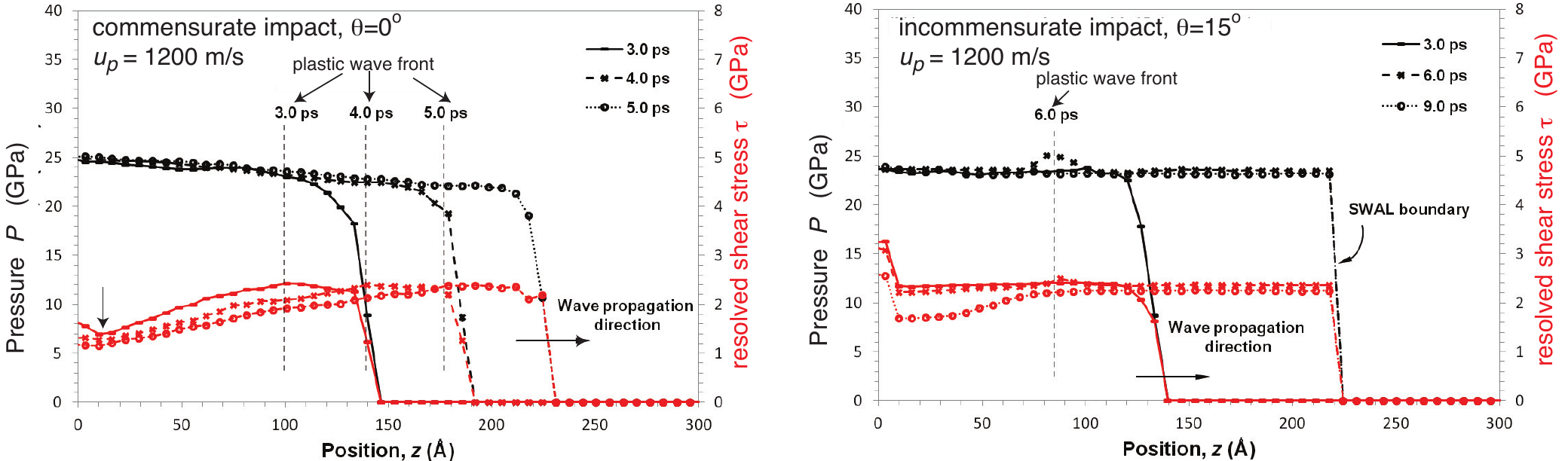}%
\caption{Pressure and resolved shear stress profiles for (a) commensurate and (b) $\theta=15^\circ$ incommensurate impact, for impact velocity $u_p =1200$\,m/s. Gray dashed line shows the position of plastic wave front. For clarity, half of the profiles are shown.}%
\label{fig:figure6}
\end{center}
\end{figure}

The evolution of the shock wave profiles yields insight into the nature of plastic deformation. Figure~\ref{fig:figure6}a shows the spatial variation of the pressure and the resolved shear stress away from the interface at three different simulation times for a relatively high velocity impact, $u_p =1200$\,m/s. The glide occurs along $\{111\}$ close-packed planes with the maximum Schmid Factor (SF$^{m}$). In each case, the shock wave splits into an elastic-plastic two-wave front. This is also evident in the thermo-mechanical contour plots shown in Fig.~\ref{fig:figure3}. The elastic precursor that separates the leading elastic wave front from the plastic front where the resolved shear stress is a maximum (dashed lines) is a few nanometers thick. The trends in the stress profiles, away from the interface and within the shockwave are strikingly different from those observed in studies on bulk crystals~\cite{swp:HolianLomdahl:1998, swp:RaveloHolian:2014}. 

At the interface, the shear stress is non-zero. It evidently exceeds the threshold for dislocation nucleation since the interface continues to operate as a source (Fig.~\ref{fig:figure4}b).  Away from the interface, there is a marginal, monotonic decrease in the pressure right to the plastic zone. Past the plastic wave front and within the elastic precursor, there is a more rapid decrease. The shear stress gradually increases away from the interface. It reaches a maximum at the plastic front, stabilizes within the elastic precusor and then discontinuously decreases at the elastic front. Within the plastic zone, the shear stress decreases in time due to the intersection of the Shockley partials away from the interface. The stress distribution is unlike that observed in bulk shockwaves in that the discontinuous decrease in the pressure and the corresponding rapid increase in the shear stress at the plastic front are both absent~\cite{swp:HolianLomdahl:1998}. 

The variations are directly attributable to interfacial plasticity and the associated relaxation processes. In particular, the profile is modified by partial dislocations that nucleate at the interface and expand into the bulk. These events delay the onset of the plastic zone, resulting in decrease in dislocation density away from the interface compared to that in the bulk studies, eventually leading to a modified stress profile at the plastic wave front. 

In some cases, especially during early stages of the impact, there is a rapid decrease in the shear stress over a few interatomic spacings away from the interface region. This is evident in the $3$\,ps profile in Fig.~\ref{fig:figure6}a. To see why, note that the nucleation of the partials from the interface region following impact increases the dislocation content at the interface as the Burgers vector of the entire bicrystal system must be conserved\footnote{This is only true if we ignore non-conservative structural transitions at the interface. At low temperatures, this is a reasonable approximation, although it may breakdown at large impact velocities where the interface temperature can approach the melting point.}.  As an example, the interface configuration in Figure~\ref{fig:figure4}b-4 consists of remnants of the trailing partials following dislocation nucleation at multiple interfacial sites. The large strain accommodation occurs preferentially in the plane of the defective interface, and elevates the shear stress compared to that in the crystalline surrounding bulk. The effect should be more pronounced for incommensurate interfaces, and this is indeed the case as we show in detail below.

\subsubsection{Incommensurate impact} 
\begin{figure}
\begin{center}
\includegraphics[width=\columnwidth]{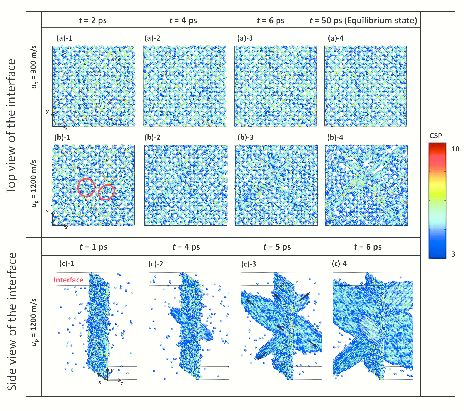}%
\caption{Same as Fig.~\ref{fig:figure4}, but for impact between incommensurate surfaces misoriented by $\theta=15^\circ$ about the $[001]$-axis, for two impact velocities, $u_p=300$ and $1200$\,m/s.}
\label{fig:figure7}
\end{center}
\end{figure}

Figure~\ref{fig:figure7} shows the atomic configurations around the interface for incommensurate impact between surfaces twisted by $\theta=15^\circ$ about the [001]-axis. The impact results in the formation of a twist grain boundary (TGB) at $z=0$. Although the TGB consists of significant number of non-FCC atoms as identified by the CSP parameter, the interface region is structured. It is associated with a primitive supercell based on the coincident lattice sites at the interface. The integer ratio of this primitive cell to the lattice primitive cell $\Sigma$ is often used as a measure of the interface structure; smaller the $\Sigma$ value, more structured the grain boundary~\cite{gbe:SuttonBalluffi:1987, gbe:Randle:1999}. The $\theta=15^\circ$ $[001]$ TGB is close to the high symmetry $\Sigma 25 (\theta=16.3^\circ)$ boundary. At low velocities ($u_p = 300$\,m/s, Fig.~\ref{fig:figure7}a), the interface structure is stable with no point or extended defects, i.e. the interface is able to absorb the entire impact elastically and the velocity is sub-critical.

Systematically increasing the impact velocity yields the critical point associated with incipient plasticity, $u_p^\ast\approx1200$\,m/s. This value is well above that for the commensurate response, yet the extent of plasticity is much higher, evident from the rapid increase in the number density of HCP atoms at $u_p^\ast\approx1200$\,m/s (Fig.~\ref{fig:figure5}). The corresponding atomic configurations at the interface and within the bulk are shown in Fig.~\ref{fig:figure7}b-c. During the initial stages, we see nucleation of localized defective regions at the interface. They are composed of point defects, mainly vacancy-interstitial pairs  (circled, Fig.~\ref{fig:figure7}b-1). These defects serve as precursors to plastic yielding at the interface. After a critical incubation time, one of the defective regions leads to the nucleation and glide of a pair of partial dislocations into the adjoining bulk lattices (Fig.~\ref{fig:figure7}b-2). This is followed by nucleation of two additional pairs of dislocations such that all four $\{111\}$ close-packed planes are activated from the defective region (Fig.~\ref{fig:figure7}b-3). Eventually, we see slip on parallel glide planes from adjoining defective regions (Fig.~\ref{fig:figure7}b-4).

Although the four slip planes have the same SF$^m$, the fact that they are not activated simultaneously indicates that the interface structure plays a crucial role in setting the competition between the slip planes. Past studies have shown that structure and distribution of morphological features such as kinks and steps at the grain boundaries can lead to dissociation of grain boundary dislocations (GBDs) along preferred directions into the adjoining crystals~\cite{gb:TschoppMcDowell:2008a, gb:TschoppMcDowell:2008b, gb:DerletWang:2009}. In some cases, the dissociation can occur along low SF planes. The subsequent plasticity is then biased by the presence of these dissociated dislocations. For example, dislocation nucleation from the boundary usually increases its dislocation content. However, if the stress accommodation occurs via growth of the pre-dissociated dislocations, the reverse is true as the boundary loses part of the dissociated dislocation. In the case of [001] TGBs studied here, the boundary is composed of screw dislocations and steps that are sensitive to the degree of incommensuration or twist misorientation, and the boundary inclination is singular~\cite{book:SuttonBalluffi:1995}. The structural features will modify the nucleation of the leading Shockley partials along preferred planes. We should emphasize that in the case of high velocity impact, the processes occur within a TGB structure that is itself in the process of attaining its equilibrium structure, a combined effect of an elevated stress state and temperature. Although the evolving interfacial structure is still expected to modify the mechanistic details of dislocation nucleation, they will naturally be modified compared to the response of an equilibrated grain boundary. For example, the unaxially compressed state following release of the shockwaves can suppress the dissociation of grain boundary dislocations. The fact that the growth of the leading partials occurs along high SF close-packed planes is suggestive that this is the case. 
As additional confirmation, in time the accumulation of dislocation content within the TGB results in structural breakdown (Fig.~\ref{fig:figure7}b-4). We see evidence of inhomogeneous strain distribution, and enhanced shear stress at the GB relative to the surrounding bulk (Fig.~\ref{fig:figure6}b).

The critical velocity remains unchanged for the $\theta=30^\circ$ (near $\Sigma 17 - \theta=28.1^\circ$ boundary) TGB, yet the incipient plasticity is different in that it is limited to slip on parallel $(1\bar{1}1)$ planes.  The change in TGB structure due to the increase in misorientation suppresses the activation of the other three close-packed planes. Since the nucleated leading partials do not intersect as they grow into the bulk, there is no loss of plastic strain energy away from the interface. The number density of HCP atoms again increases rapidly at $u_p=1200$\,m/s (Fig.~\ref{fig:figure5}). The increase is larger than that for $\theta=15^\circ$, indicating that the overall plasticity is enhanced despite the anisotropic activation of one of the four slip planes. Further increasing the twist to $\theta=45^\circ$ (near $\Sigma 29 - \theta=43.6^\circ$ boundary) results in a TGB structure that is able to absorb the $u_p=1200$\,m/s impact elastically, i.e. the response to impact is sub-critical.  

The yield point associated with the impact is also sensitive to the interface type. For both $\theta=15^\circ$ and $\theta=30^\circ$ TGBs, $\sigma_Y\approx24$\,GPa. The value is more than twice higher than that for the commensurate case. The increase in yield point correlates with an increasing incubation time for the TGBs to release the plastic deformation following impact. As an example, the shear profiles for the $\theta=15^\circ$ impact shown in Fig.~\ref{fig:figure6}b are associated with a significantly wider elastic precursor. The  3\,ps profile reveals a  well-formed elastic front but the plastic front is absent; it has not yet nucleated from the TGB. At 6\,ps, the plastic front has nucleated and moved several interatomic spacings away from the TGB, but the elastic front has moved past the SWAL. Since the pressure $P$ does not change significantly with $\theta$, the larger width implies that the elastic energy stored in the precursor is larger.

\subsection{Thermo-mechanical analysis}
\label{sec:tmanalysis}
\subsubsection{Temperature profiles}
\begin{figure}
\begin{center}
\includegraphics[width=\columnwidth]{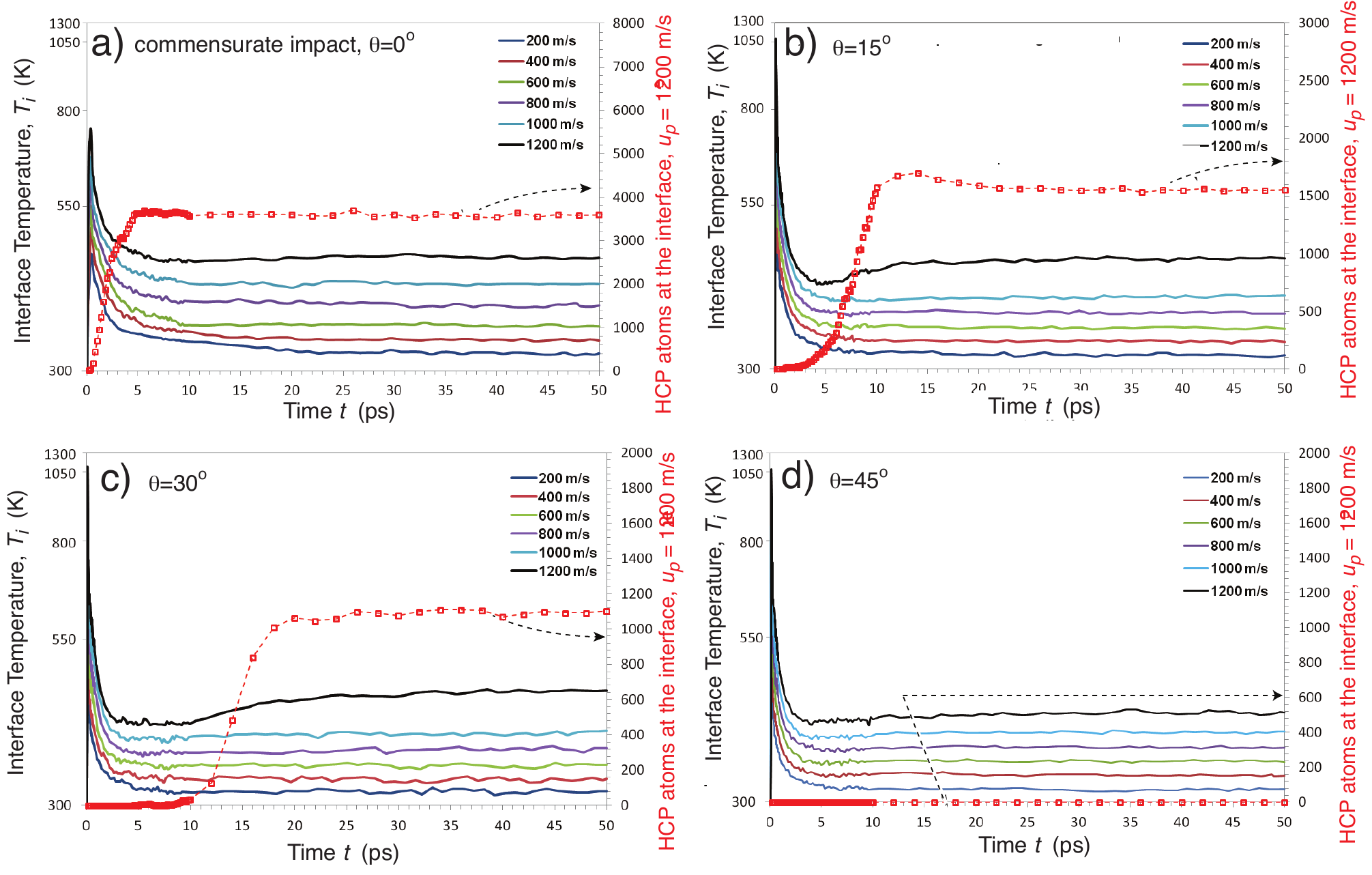}%
\caption{The temporal of the interfacial temperature $T_i(t)$ for varying impact velocities and degree of commensuration. The temperature is plotted on a logarithmic scale. The plots also show the evolution of HCP atoms within the interface for impact velocity $u_p=1200$\,m/s.}%
\label{fig:figure8}
\end{center}
\end{figure}
Figure~\ref{fig:figure8} shows the temporal evolution of average interface temperature $T_i(t)$ for varying impact velocities and degree of commensuration. In almost all of the cases, the initial kinetic energy transfer to the interface atoms results in a rapid increase in temperature to its maximum value, followed by a much slower decay to thermal equilibrium. The maximum temperature occurs on the order of $0.1$\,ps. The time-scale is of the order of lattice vibrations in elemental metals and relatively insensitive to the impact velocity and the degree of incommensuration, indicating that the initial temperature rise is set by the post-impact displacive phonon emission within the interface. 

The slower decay to equilibrium takes place over 5-10\,ps. At and above the critical impact velocity, the interface begins to operate as a source of dislocations over this time-scale. Careful examination of the atomistic processes indicates that this part of the thermal evolution is associated with redistribution of the interfacial atoms (Fig.~\ref{fig:figure4}b). To better quantify their extent, we monitor the HCP atoms generated within the interface region. The plots in Fig.~\ref{fig:figure8} show their temporal evolution for $u_p=1200$\,m/s. The commensurate impact is super-critical and HCP atoms nucleate at impact ($t=0$). Their number density increases rapidly past the maximum temperature and saturates as the temperature starts to equilibrate ($t\approx5$\,ps). The interfacial response following the $\theta=15^\circ$ and $\theta=30^\circ$ incommensurate impacts is surprising in that the temperature does not decay monotonically. Rather, the initial decay is followed by an even slower increase to an equilibrium temperature over $\approx20$\,ps. The impact is critical, and as discussed before, the incipient plasticity is delayed due to structural redistribution at the interface. The resultant incubation time registers as a lag in the nucleation of HCP atoms at the interface and is correlated with the slow increase in temperature, indicating that the additional increase in temperature is due to plastic deformation at and away from the interface. The delay in nucleation of HCP atoms increases with $\theta$, either due to decreasing interface energy with misorientation or enhanced ability to absorb the impact via interface structural reconstructions. Eitherway, the trend again underscores the effect of interfacial structure on the thermo-mechanical response following impact. 

\begin{figure}
\begin{center}
\includegraphics[width=\columnwidth]{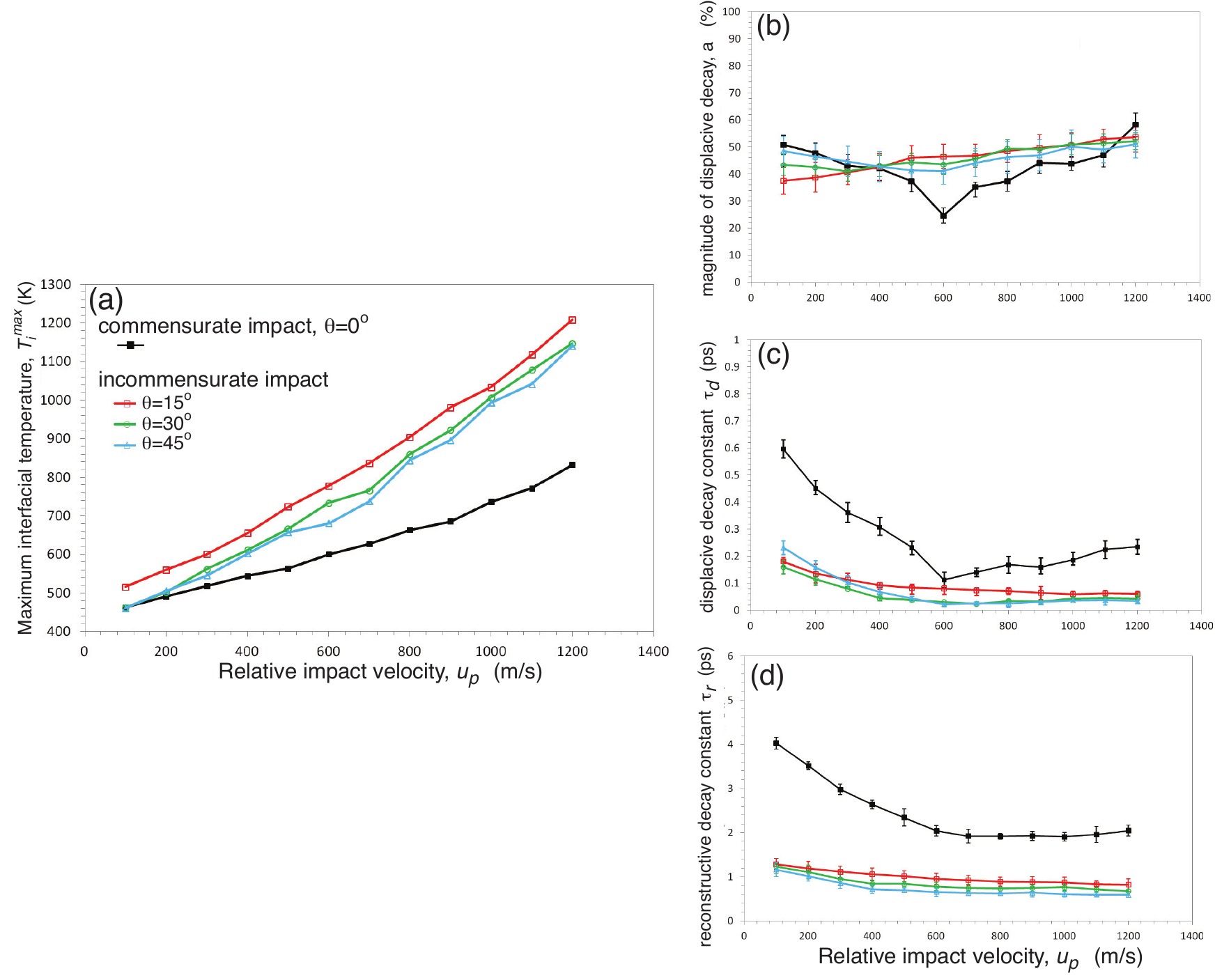}%
\caption{(a) The maximum temperature $T_i^{max}$ at the interface as a function of impact velocity $u_p$. Each data point is time-averaged over $\pm0.2$\,ps. (b-d) The dependence of the temperature decay on the impact velocity, quantified by three coefficients ($a$, $\tau_d$ and $\tau_r$) associated with the bi-exponential fit (Eq.~\ref{eq:biExp}); see text for details. Error bars represent 95\% confidence bounds of fitting coefficients.}%
\label{fig:figure9}
\end{center}
\end{figure}

The maximum interfacial temperature $T_i^{max}$ associated with the temperature profile varies significantly with the impact velocity and interface type, and the dependence is shown in Fig.~\ref{fig:figure9}. In all cases that we have explored, $T_i^{max}$ is well below the bulk or interfacial melting point~\cite{premelt:DahmenJohnson:2004, premelt:FensinHoyt:2010, fec:HoytUpmanyu:2010}. For commensurate impact, the sub-critical increase in $T_i^{max}$  is roughly linear with a slope $\approx0.1$\,K/(m/s), and it becomes weakly super-linear above the critical velocity. The $T_i^{max}$ for incommensurate impact is always higher. For sub-critical impact, this is not surprising as the interface is able to absorb and dissipate part of the kinetic energy during the adiabatic compression, and the displacive phonons are absorbed preferentially within the relatively open interfacial structure. The increase in $T_i^{max}(u_p)$ is super-linear and more pronounced compared to the commensurate case. The maximum temperature decreases with degree of incommensuration, a reflection of their relative structural robustness and/or reconfigurability during the impact. The trends are likely modified due to the variations in the phonon mean free path within the structured TGBs, as well as the ability of the TGBs to undergo non-equilibrium structural transitions~\cite{gb:VitekSuttonSchwartz:1983}. 

In order to quantify trends in the temperature decay, the relevant portion of the temperature profile is fit to a biexponential function, 
\begin{equation}
\label{eq:biExp}
T_i (t)= T_i^{eq} + A\left[ ae^{-t/\tau_d} + (1-a)e^{-t/\tau_r}\right].
\end{equation}
In each case\footnote{The $\theta=15^\circ$ and $\theta=30^\circ$ cases exhibit an additional temperature increase at $u_p = 1200$\,m/s and this is eliminated by limiting the fit to $\approx8$\,ps.}, the coefficient of determination $R^2=0.98-0.99$, indicating that the temperature decay is well-described by Eq.~\ref{eq:biExp}. The function is physically motivated as the overall decrease is the sum of two underlying processes, a fast decay associated with recovery of the displacive phonons, and slower recovery due to interfacial reconstructions. The parameter $a$ is the weight that partitions the response between these two interfacial processes, and $\tau_d$ and $\tau_r$ are the time constants associated with the displacive and reconstructive recovery, respectively.

The fitting coefficients are sensitive to $u_p$ and $\theta$, and their dependence is plotted in Fig.~\ref{fig:figure9}b-d. The weight parameter $a$ is the magnitude of the displacive contributions to the decay, and for the commensurate impact it is a minimum at the critical impact velocity, $u_p^{\ast}=600$\,m/s, i.e. the temperature decay is dominated by structural reconstructions in the vicinity of the  critical impact as the interface destabilizes and then structurally heals following defect nucleation. Further increasing the velocity enhances the displacive contributions as the interfacial processes that lead to plastic deformation saturate. The trends for incommensurate impacts are qualitatively different. Below the critical velocity, the contribution of the two types of processes is approximately equal ($a\approx50\%$), although at low velocities the displacive contribution increases with misorientation. The trend indicates increasing structural reconfigurability of the TGBs with misorientation. At the critical impact for $\theta=15^\circ$ and $\theta=30^\circ$, the difference in the response of the TGBs is minimal, and unlike the commensurate case we do not observed a significant decrease in the magnitude of $a$, i.e. the temperature decay is less sensitive to the interface type. 

The decay time constants associated with the two contributions yield additional insight into the dynamics of the temperature evolution. For commensurate impact, the short time displacive decay constant is of the order of $\tau_d=0.1-0.6$\,ps, consistent with the typical phonon emissive times in elemental metals. It shows a pronounced minima at the critical velocity as the plastic deformation at the interface distorts the interfacial structure and decreases the phonon contribution to heat dissipation within the modified interfacial structure. The slower reconstructive recovery occurs with decay constants in the range  $\tau_r=1.9-4.0$\,ps, and their value decreases monotonically upto the critical velocity.. Both decay constants are lower in magnitude for incommensurate impact ($\tau_d=0.02-0.2$\,ps and $\tau_r= 0.6-1.2$\,ps) with no noticeable differences in the response of the TGBs. 
\begin{figure}
\begin{center}
\includegraphics[width=0.75\columnwidth]{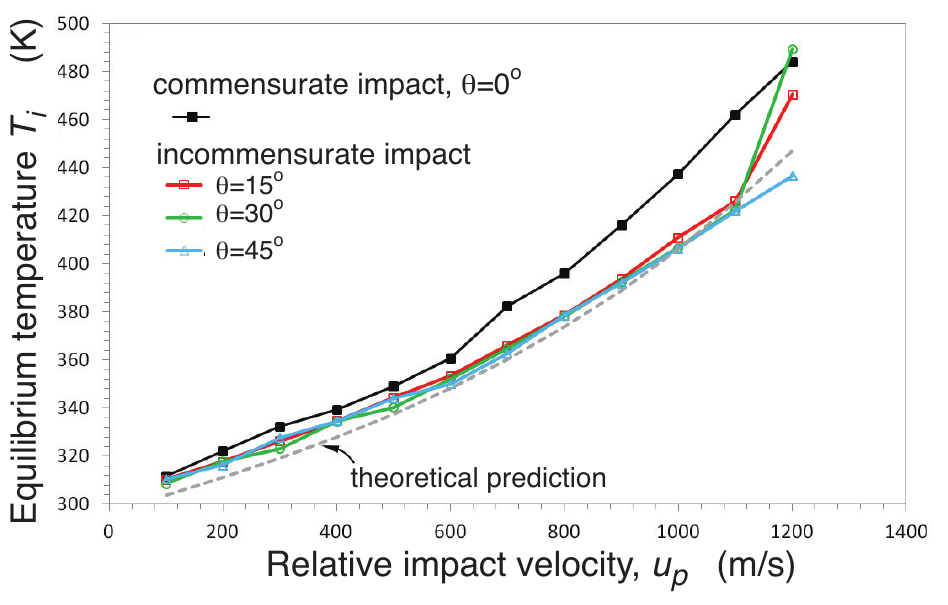}%
\caption{The equilibrium temperature $T_i^{eq}$ at the interface as a function of impact velocity $u_p$. The data points are averaged within a $2$\,ps relaxation after the entire $50$\,ps run.}%
\label{fig:figure10}
\end{center}
\end{figure}

In all cases, the interfacial temperature following the 50\,ps simulations is same as in the bulk (Fig.~\ref{fig:figure3}b), i.e. the time-scale is sufficient for the interface to attain its equilibrium temperature $T_i^{eq}$. The equilibrium temperature increases non-linearly with impact velocity (Fig.~\ref{fig:figure10}). In order to quantify the effect of the interface-mediated processes, we have also plotted adiabatic temperature increase in shock compressed  bulk solid $\Delta T_H$ based on the Gr\"uneisen equation of state~\cite{book:Meyers:1994},
\begin{align}
\Delta T_H =& T_0\, e^{\Gamma_0\left(1-\frac{V}{V_0}\right)} + \frac{V_0-V}{2C} P \nonumber\\
&+ \frac{e^{\frac{T_0V}{V_0}}}{2C}\, \int_{V_0}^{V} P\,e^{\frac{T_0V}{V_0}}\left[2-\Gamma_0\left(1-\frac{V}{V_0}\right)\right] dV,
\end{align}
where $T_0$ and $V_0$ are the initial temperature and specific volume respectively, $V$ is the compressed specific volume, $P$ is applied shock pressure, $C$ is the specific heat and $\Gamma_0$ is the Gr\"uneisen constant that can be expressed in terms of linear Hugoniot relationship, as reported recently by Bringa et al~\cite{swp:BringaCaturla:2004}. The theoretical curve based on the simulation parameters is also plotted in Fig.~\ref{fig:figure9}. Below the critical velocity, the predicted behavior consistently underestimates the equilibrium temperature, although the response is qualitatively similar to that extracted from the simulations. The difference is the order of $2-4$\,K and serves as a quantitative measure of the effect of the interface. The difference increases markedly above the critical velocity due to interface-mediated plastic deformation. 

The equilibrium temperature is also sensitive to the interface type. The temperature following incommensurate impact is almost always lower as the TGBs are able to absorb part of the impact energy via structural reconstructions. For the TGBs studied here, the decrease is of the order of $2-4$\,K. In the case of $\theta=15^\circ$ and $\theta=30^\circ$, we see a rapid increase in $T_i^{eq}$ at the critical impact velocity ($u_p^\ast=1200$\,m/s). A similar increase is also observed for the commensurate case ($u_p^\ast=600$\,m/s), although the increase is less pronounced. The incipient plasticity at these impact velocities results in defective interfacial structures that hinder the ability of the interfaces to absorb the displacive phonos and undergo structural transitions. The combined effect results in an increase in the equilibrium temperature.

\subsubsection{Stress profiles} 
\begin{figure}
\begin{center}
\includegraphics[width=\columnwidth]{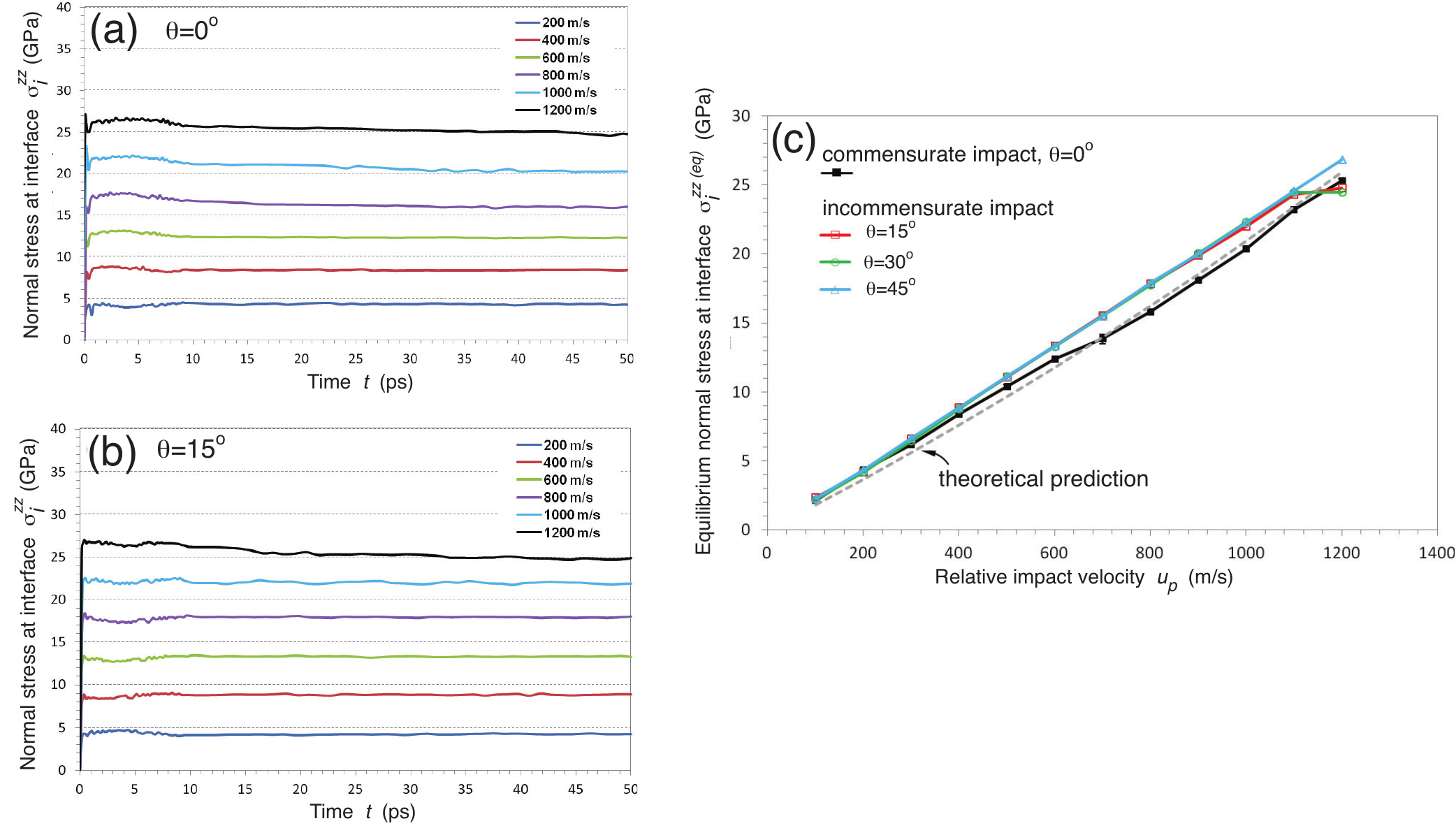}%
\caption{(a-b) Temporal evolution of the axial virial stress $\sigma_i^{zz}$ at the interface for (a) commensurate and (b) $\theta=15^\circ$ incommensurate impact. (c) The steady-state equilibrium axial stress $\sigma_i^{zz(eq)}$ as a function of impact velocity for commensurate and incommensurate impacts. The theoretical $P-u_p$ curve based on the bulk response of a Gr\"ueisen solid (Eq.~\ref{eq:Pup}) is also shown for comparison (dashed curve).}%
\label{fig:figure11}
\end{center}
\end{figure}

Figure~\ref{fig:figure11}a-b shows the temporal evolution of the virial stress component along the interface normal $\sigma_i^{zz}(t)$ as a function of impact velocity for commensurate and $\theta=15^\circ$ impacts. As expected, the normal stress increases with $u_p$. It rises on impact to a maximum over $\approx1$\,ps, and as in the temperature profile, it is of the order of the phonon emissive time-scales. Below the critical velocity, the stress profile plateaus to an equilibrium value, and the behavior is similar to that observed in bulk shock wave studies~\cite{swp:BringaCaturla:2004, swp:CaoBringaMeyers:2007}, i.e. the interfaces have little effect. Above the critical velocity, we see a sharp spike to a maximum over a much shorter time-scale ($\approx0.1$\,ps). The initial spike is associated with a uniaxially compressed state that is unstable and relaxes to a hydrostatic compression via defect nucleation. Eventually, the hydrostatic stress state undergoes a non-monotonic decay to an equilibrium value. Recent experimental X-ray diffraction studies by Higginbotham et al.~\cite{swp:HigginbothamWark:2012} have shown that dislocations in crystalline copper result in enhanced hydrostatic compression, and these findings are in agreement with our results. The effect of the interface type is minimal, although we see a relatively slower decay to the equilibrium value at the critical velocity, evident in the $\theta=15^\circ$ TGB stress profile (Fig.~\ref{fig:figure11}b). The response is consistent with slower structural reconstructions following dislocation nucleation and the increase in Burgers vector at the interface. 

The steady-state equilibrium normal virial stress $\sigma_i^{zz(eq)}$ increases almost linearly with impact velocity (Fig.~\ref{fig:figure11}c). For $\theta=0$, below the critical velocity the behavior is well-described by the bulk response of a Gr\"uneisen crystal~\cite{book:Meyers:1994}, 
\begin{align}
\label{eq:Pup}
P=\frac{\rho_0}{2}\left(v_s u_p + \frac{\Gamma_0 u_p^2}{2}\right)\,,
\end{align}
where $\rho_0$ is the initial specific density and $v_s$ is the bulk speed of sound. As expected, the theoretical prediction is lower for the TGBs as the interface structure modifies the stress state. At and above the critical velocity, there is a noticeable decrease in the normal stress, in particular for the incommensurate $\theta=15^\circ$ and $\theta=30^\circ$ TGBs as the reconstructive events at the interface following defect nucleation modify the transition from uniaxial to hydrostatic compression.

\begin{figure}
\begin{center}
\includegraphics[width=\columnwidth]{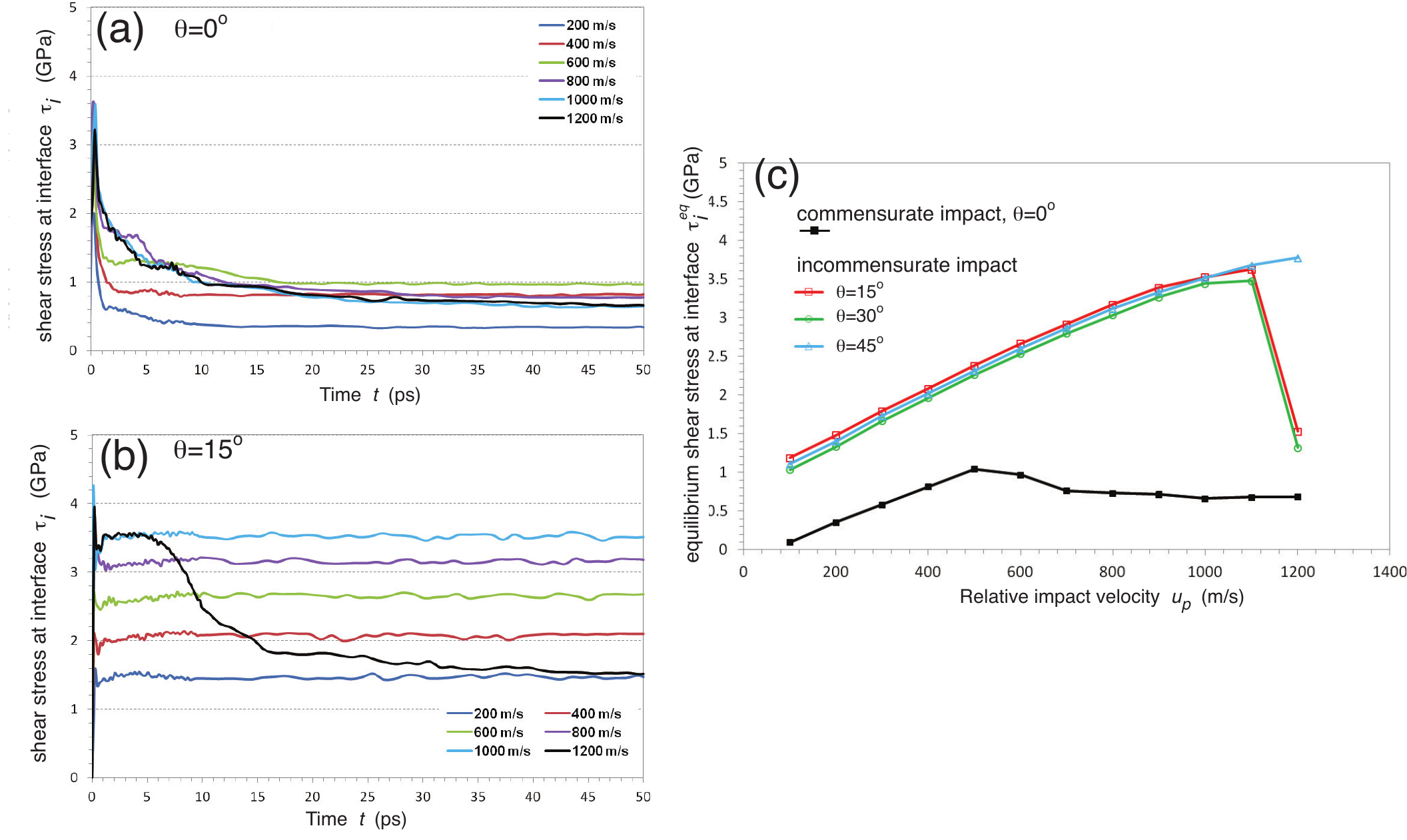}%
\caption{(a-b) Same as in Fig.~\ref{fig:figure11}, but for the resolved shear stress at the interface $\tau_i$. (c) The steady-state equilibrium shear stress $\tau_i^{eq}$ as a function of impact velocity for commensurate and incommensurate impacts.}%
\label{fig:figure12}
\end{center}
\end{figure}

The profiles of the interfacial resolved shear stress $\tau_i(t)$ are plotted in Fig.~\ref{fig:figure12}. Clearly, the evolution is sensitive to the interface type. Below the critical velocity, the commensurate response is similar to that observed for interface temperature and normal stress (Fig.~\ref{fig:figure12}a); the shear stress increases rapidly to a maximum and then decays slowly to a steady-state equilibrium value $\tau_i^{eq}$. Unlike the normal stress, the decay is considerably slower and larger in extent compared to the  incommensurate response (Fig.~\ref{fig:figure12}b). As a result, the equilibrium shear stress $\tau_i^{eq}$ for commensurate impact is lower (Fig.~\ref{fig:figure12}c), indicating that the energy dissipation at the interface and the associated interface reconstructions facilitate absorption of the generated shear stress. At and above the critical impact, the equilibrium shear stress is lowered by the generation and annihilation of interfacial defects and dislocation nucleation. The decrease is more dramatic for incommensurate interfaces ($\theta=15^\circ$ and $\theta=30^\circ$ TGBs), again highlighting the role of interface structure.

\subsubsection{Interfacial energetics} 
Figure~\ref{fig:figure13}a shows the effect of the impact velocity on the interface energy, as defined by Eq.~\ref{eq:intEnergy}. The quantitative trends at low velocities are close to their equilibrium values as the impact related changes in interface structures are minimal. For commensurate impact, the interface energy $\gamma_i^H\approx0.05-0.1$ J/m$^2$ and is a reflection of the the remnant point defects following the reconstructive events. It decreases continuously to zero, as expected for a perfect crystal. It is higher for incommensurate interfaces ($\gamma_i^H\approx0.4-0.7$ J/m$^2$), and increases with misorientation $\theta$. The values are consistent with past reports on energies (enthalpies) of grain boundaries in FCC metals~\cite{gbe:SuttonBalluffi:1987, gbe:NajafabadiSrolovitzLeSar:1991, gbe:ErnstFinnis:1996, gbm:Zhang:2005} Discounting the entropic effects, the trend reflect an increase in the energy with misorientation of the TGBs. Interestingly, the $\theta=45^\circ$ TGB that has the highest energy is also associated with the largest critical velocity, suggesting that higher energy interfaces have better impact resistance, presumably due to lower barriers to surface reconstructions.

Below the critical impact, increasing the impact velocity results in non-monotonic variations in the enthalpies of the TGBs. The interface energy initially increases with $u_p$, and reaches a maximum, and then decreases just prior to the critical impact. As discussed before, the trends are not surprising as they reflect the interplay between the non-equilibrium thermo-mechanical state of the interfaces and the surrounding bulk, and their respective energies, $\gamma_i^H\equiv\gamma_i^H(T, \sigma_{\alpha\beta})$. The initial increase is likely due to structural reconstructions at the interfaces as the deformation of the bulk is minimal. At larger velocities, though, the higher temperature at the interface and the increasingly (elastically) deformed state of the bulk following the SW absorption becomes important. The combined effect lowers the excess energy at the interface, and therefore its energy. 

At the critical velocity, the nucleation of defects and dislocations at the interfaces increase their energy. This increase is evident for all interfaces. It is  more substantial for the $\theta=15^\circ$ and $\theta=30^\circ$ TGBs, indicating a higher sensitivity to the presence of non-equilibrium defects and extrinsic dislocation content. The interface energies of these two TGBs $\gamma_i^H\approx1.2$\, J/m$^2$, an almost two-fold increase over the near-equilibrium value at  $u_p=100$\,m/s. For the commensurate case, the energy continues to increase non-linearly for $u_p>u_p^\ast$. At $u_p=1000$\,m/s, the energy is almost an order of magnitude higher than that at $u_p=100$\,m/s.

\subsubsection{Work of adhesion}
Our results show that the dissipation of the input kinetic energy $E_{KE}$ can be significantly underestimated by ignoring interfacial structure, plasticity and energetics.  The dissipation is crucial for the stability of the metallurgical bond formed post impact and traditionally the elasto-plastic deformation within the bulk is assumed to be the dominant dissipation mechanism. The thermo-mechanical state of the interface and the bulk following impact is critical as the non-equilibrium interface enthalpies (and likely their free energies) are significantly higher. These non-equilibrium effects serve to reduce the available elastic energy following impact $E_{KE}-E_p$, where $E_p$ is the sum of total of energy dissipated via plastic deformation in the bulk. In the extreme case that the strength of the metallurgical bond or work of adhesion $W$ between the impacting surfaces is less than the available elastic energy $E_{KE}-E_p>W$, the two surfaces will not bond following impact. 

\begin{figure}
\begin{center}
\includegraphics[width=\columnwidth]{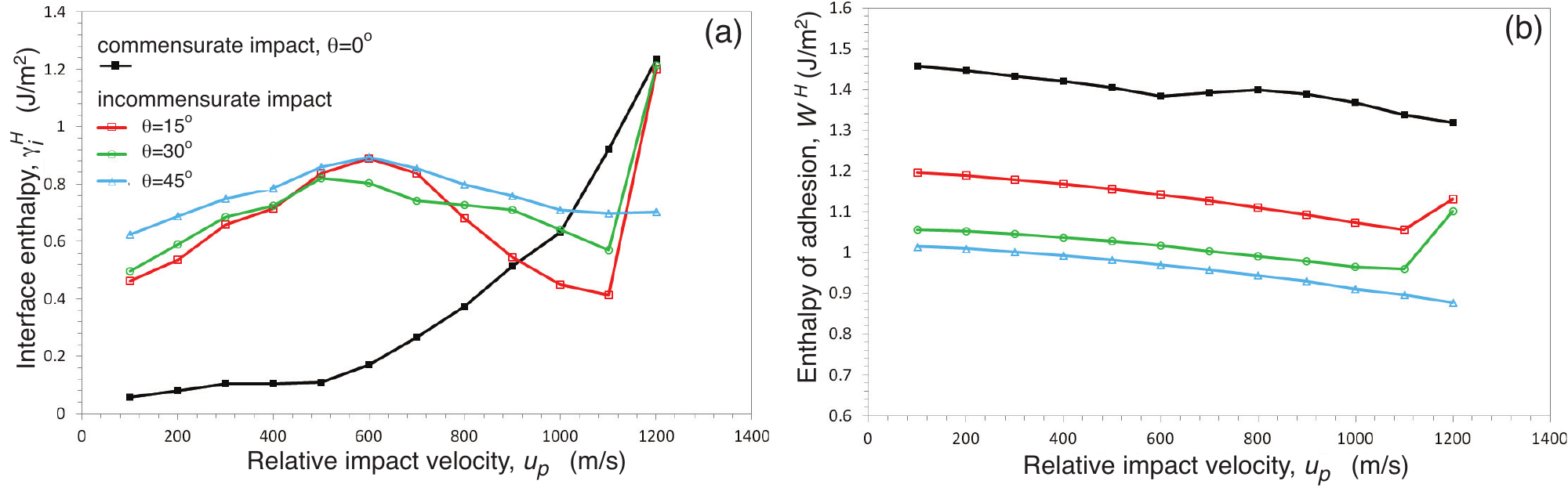}%
\caption{(a) Interface energy $\gamma_i^H$ and (b) enthalpy of adhesion $W^H$ as a function of impact velocity. Each data point represents an average over a $2$\,ps relaxation run following the $50$\,ps simulation.}%
\label{fig:figure13}
\end{center}
\end{figure}

Predicting the bond strength is obviously important for several key high strain-rate processes. We quantify the effect of non-equilibrium energetics and kinetics by extracting the enthalpic component of the work of adhesion, $W^H$ (Eq.~\ref{eq:WOA}). The quantity is a dynamic measure of the excess energy stored at the interface with respect to the energies (surface enthalpies) of the impacting surfaces. Note that the surface enthalpy is extracted by freezing the atoms in the vicinity of the interface following deformation, as it represents the self-consistent reference for the interfacial strength. The effect of the impact velocity is plotted in Fig.~\ref{fig:figure13}b. The low impact velocity response shows that the commensurate impact has the higher work of adhesion, as expected, and it decreases with increasing $\theta$ for incommensurate interfaces, i.e. interfaces with higher enthalpies result in lower adhesion, $W^H\propto 1/\gamma_i^H$. Increasing the impact velocity results in a linear decrease in $W_i^H$ for all interfaces; the sum total of non-equilibrium effects is to always lower the strength of the metallurgical bond between the interfaces. At the critical velocity, there is again an increase in $W^H$ as the processes associated with defect and dislocation nucleation dissipate significant energy into the bulk. Above the critical velocity, there is a slight decrease in $W^H$ for the commensurate case, possibly due to the competing effects of enhanced defect nucleation and their annihilation by plastic deformation. 

\section{Summary}
Our computations on impact-induced plasticity of $[001]$ copper surfaces highlight the role of the structure of the interface formed post-impact on the subsequent thermo-mechanical processes and bond formation. The impact-induced plasticity is mediated by shock waves within both crystals that split into two wave fronts as they propagate through the crystals, resulting in a gradual decrease in the pressure and a concomitant increase in resolved shear stress away from the interface. The 1D uniaxial compressive stress within the interface zone increases linearly with impact velocity below yield point.  Above a threshold velocity, this uniaxial compressed state is relaxed via hydrostatic compression at the interface. Past the critical point, the interface serves as a nucleation site for non-equilibrium point defects such as vacancy clusters, and dislocation nucleation and emission. The structure of the post-impact interface has a significant effect on the critical point for shock-induced incipient plasticity; the yield point is $\sigma_Y\approx11$\,GPa for the commensurate impact (twist misorientation $\theta=0$)  and is much lower than that the bulk value of $32$\,GPa~\cite{swp:CaoBringaMeyers:2007}. It increases to $\sigma_Y=24$\,GPa for $\theta=15^\circ$ and $\theta=30^\circ$ incommensurate impacts, while no plasticity is observed for the $\theta=45^\circ$ impact over the 50\,ps simulated time interval. 

The interfacial plasticity is associated with non-equilibrium structural modifications due to nucleation of point defect clusters (such as vacancy clusters) within the structural units. Below the critical point, the interfacial energy increases with increasing impact velocity due to structural reconstructions necessary to accommodate the point defects. Above yield point, the interface energy is further modified by point defect/dislocation traces following their nucleation, and the temperature increase at the interface. In the case of incommensurate interfaces such as the twist grain boundaries studied here, significant plasticity is absorbed by the interfacial structure and we see evidence of cases where the interface energy {\it decreases} for a range of super-critical impact velocities. The  interfacial structure also modifies the resolved shear stress; it increases with the degree of incommensuration between the crystals. The extent of material exchange greatly increases followed the growth of stacking faults associated with the Shockley partials nucleating from the interfacial zone.

The temperature increase is particularly important for understanding the thermo-mechanical response, and also as input for multi-scale computations aimed at understanding the impact at experimental scales. To this end, we also quantify the nanoscale phenomena that dictate the phonon transport-based temperature increase within the interface zone\footnote{The electronic contribution to thermal transport is ignored within the classical computations, and this may be reasonable as high impact velocities should result in phonon-dominated thermo-mechanical processes.}. These include elastic isentropic or adiabatic work, plastic work mediated by dislocation nucleation and their propagation within the crystals, and heat absorption at the interface. For the range of impact velocities employed in this study ($100-1200$\,m/s), the impact temperature does not reach the melting point of copper. Again, the non-equilibrium interface structure plays a non-trivial role in modifying the temperature increase, as a significant portion of the phonon energy is dissipated efficiently within the structural units that make up the interface. Specifically, plasticity induced heating in this work is high, $\sim30-60$\,K increase for every $100$\,m/s increment in the impact velocity.  

Our results have direct relevance for micron-scale particle impact techniques such as cold spray where a material-specific critical velocity for successful metallic bonding is often reported~\cite{swp:KurodaKatanoda:2008}. It is interesting to compare our simulation results with the cold spray experimental observations, where past reports indicate that the critical velocity for copper on copper impact lies in the range 340-600 m/s~\cite{swp:KurodaKatanoda:2008, swp:AssadiKreye:2003}. Although out study is limited to single crystalline copper crystals and ignores the effect of pre-existing defects as well as the asymmetry in the momenta between the particle and the substrate, our results imply that a strong interfacial structure is possible in this velocity range for incommensurate interfaces, and serves as a potential explanation for metallurgical bond formation during cold spray related processes.

{\bf Acknowledgements}: This work is supported by Northeastern University Tier-I grant. We thank Profs. T. Ando and A. Gouldstone for useful discussions. MU acknowledges partial support from National Science Foundation DMR CMMT program (Grant \#1106214), a Short Term Innovative Research (STIR) grant from Synthesis and Processing Division at Army Research Office (Program Manager David Stepp), and U.S. Army Armament Research, Development and Engineering Center (ARDEC).

\end{document}